\documentclass[12pt]{article}
\usepackage{graphicx}
\usepackage{array,dcolumn}      
\usepackage{axodraw}
\usepackage{amssymb}
\renewcommand{\thefootnote}{\arabic{footnote}}

\newcommand{\hs}{\hspace*{0.5cm}}
\newcommand{\be}{\begin{equation}}
\newcommand{\ee}{\end{equation}}
\newcommand{\bea}{\begin{eqnarray}}
\newcommand{\eea}{\end{eqnarray}}
\newcommand{\baa}{\begin{eqnarray*}}
\newcommand{\eaa}{\end{eqnarray*}}
\newcommand{\bary}{\begin{array}}
\newcommand{\eary}{\end{array}}
\newcommand{\bit}{\begin{itemize}}
\newcommand{\eit}{\end{itemize}}

\begin{document}

\begin{flushright}
{\it To appear in J.E.T.P.}
\end{flushright}
\vspace*{0.5cm}

\begin{center}
{\large \bf  SINGLE  $Z'$  PRODUCTION \\[0.3cm]
AT CLIC BASED ON $ e^-\ \gamma $  COLLISIONS}\\
\vspace*{1cm}

 {\bf $ \mbox{D. V.  Soa}^{a}$,  $ \mbox{H.
N. Long}^{b}$\footnote{On leave from Institute of Physics, NCST,
Hanoi, Vietnam},  $ \mbox{D. T. Binh}^{c}$, } and {\bf $ \mbox{D.
P. Khoi}^{c,d}$ }
\\[0.2cm]
{\it $^a$ Department of Physics, Hanoi University of Education,
 Hanoi, Vietnam}\\
{\it $^b$ Physics Division, NCTS,
National Tsing Hua University,
 Hsinchu, Taiwan}\\
 {\it $^c$ Institute of Physics, NCST, Hanoi, Vietnam}\\
 {\it $^d$ Department of Physics, Vinh University,
 Vinh,  Vietnam}\\

 \end{center}

 \vspace*{1cm}

\begin{abstract}
We analyze the potential of CLIC based  on $e-\gamma$ collisions
 to search for new $Z'$ gauge boson. Single $Z'$
 production at $e-\gamma$  colliders in two
$SU(3)_C\otimes SU(3)_L\otimes U(1)_N$ models: the minimal model
and the model with right-handed (RH) neutrinos is studied in
detail. Results show that new $Z'$ gauge bosons can be observed at
the CLIC,  and the cross sections in the model with RH neutrinos
are bigger than those in the minimal one.

\end{abstract}
\vspace*{0.5cm}

PACS number(s):12.10.-g, 12.60.-i, 13.10.+q, 14.80.-j.\\


\section{Introduction}

\hs Neutral gauge structures beyond the photon and the $Z$ boson
have long been considered as one of the best motivated extensions
of the Standard Model (SM) of electroweak interactions. They are
predicted in  many beyond  SM  models. One of them is the models
based on $SU(3)_C\otimes SU(3)_L\otimes U(1)_N$ (3 - 3 - 1) gauge
group~\cite{pp,fr,fhft,flt,hnl}. These models have some
interesting characteristics. First, the models predict three
families of quarks and leptons if the QCD asymptotic freedom is
imposed. Second, the Peccei -  Quinn symmetry naturally occurs in
these models~\cite{pal}. Finally, the characteristic of these
models is that one generation of quarks is treated differently
from two others. This could lead to a natural explanation for the
unbalancing heavy top quark,....\\
 \hs The $Z'$ gauge boson is a necessary element of the
 different models extending the SM. In general, the extra $Z'$
 boson may not couple in a universal  way. There are, however,
 strong constraints from flavour changing neutral current processes
 specifically limiting non-universal between the first two
 generations. Low limits on the mass of
 $Z'$ following from the analysis of variety
 of popular models
 are found to be in the energy intervals 500 - 2000
 GeV~\cite{gew,lv}.\\
\hs  Recently there are some arguments that the 3 - 3 - 1 models
arise naturally from the gauge theories in space time with extra
 dimensions~\cite{qui} where the scalar fields are the components in
 additional dimensions ~\cite{hil}. A few different
 versions of the 3 - 3 - 1 model have been recently proposed~\cite{col}.\\
 \hs  Recent investigations have indicated that signals of new
 gauge bosons in models may be observed at the CERN LHC~\cite{dio}
 or Next Linear Collider (NLC)~\cite{ras,ls}. In~\cite{sil},
 two of us  have considered single production of the bilepton
 and shown that with the integrated luminosity $L\simeq 9 \times
 10^4 fb^{-1}$ one expected several thousand events. In this work,
 single production of new $Z'$ gauge boson
 in the 3 - 3 - 1 models is considered. The paper is
 organized as follows. In Section 2 we give a brief review of two
models: relation among real physical bosons and constraints on
their masses. Section  3 is devoted to  single production of the
$Z'$ boson in the  $e-\gamma$ collisions. Discussions are given in
Section 4.

\section{A review of the 3 - 3 - 1 models }

\hs To frame the context, it is appropriate to briefly recall some
relevant features of two types of 3 - 3 - 1 models: the minimal
model proposed by  Pisano,  Pleitez and Frampton
(PPF)~\cite{pp,fr} and the model with RH neutrinos
(FLT)~\cite{flt,hnl}.
\subsection{The minimal 3 -- 3 -- 1 model}

\hs The model treats the leptons as  the $\mbox{SU(3)}_L$
antitriplets~\cite{pp,fr,dng}
\renewcommand{\thefootnote}{\fnsymbol{footnote}}%
\footnote{The leptons may be assigned to  a triplet as
in~\cite{pp}, however the two models are mathematically
identical.}

 \be f_{aL} = \left( \begin{array}{c}
               e_{aL}\\ -\nu_{aL}\\  (e^c)_a
               \end{array}  \right) \sim (1, \bar{3}, 0),
\label{l} \ee where $ a = 1, 2, 3$ is the generation index.

  Two of the three quark generations transform as triplets and
the third generation is treated differently. It belongs to an
antitriplet: \be Q_{iL} = \left( \begin{array}{c}
                u_{iL}\\d_{iL}\\ D_{iL}\\
                \end{array}  \right) \sim (3, 3, -\frac{1}{3}),
\label{q} \ee
\[ u_{iR}\sim (3, 1, 2/3), d_{iR}\sim (3, 1, -1/3),
D_{iR}\sim (3, 1, -1/3),\ i=1,2,\] \be
 Q_{3L} = \left( \begin{array}{c}
                 d_{3L}\\ - u_{3L}\\ T_{L}
                \end{array}  \right) \sim (3, \bar{3}, 2/3),
\ee
\[ u_{3R}\sim (3, 1, 2/3), d_{3R}\sim (3, 1, -1/3), T_{R}
\sim (3, 1, 2/3).\] The nine gauge bosons  $W^a (a = 1, 2, ...,
8)$ and $B$ of $SU(3)_L$ and $U(1)_N$ are split into four light
gauge bosons and five heavy gauge bosons after $SU(3)_L\otimes
U(1)_N$ is broken to  $U(1)_Q$. The light gauge bosons are those
of the Standard Model: the photon ($A$), $Z_1$, and $W^\pm$. The
remaining five correspond to new heavy gauge bosons $Z_2,\ Y^\pm$
and the doubly charged bileptons $X^{\pm \pm}$. They are expressed
in terms of $W^a$ and $B$ as~\cite{dng} \bea \sqrt{2}\ W^+_\mu &=&
W^1_\mu - iW^2_\mu ,
\sqrt{2}\ Y^+_\mu = W^6_\mu - iW^7_\mu ,\nonumber\\
\sqrt{2}\ X_\mu^{++} &=& W^4_\mu - iW^5_\mu. \label{idmin1} \eea
\bea A_\mu  &=& s_W  W_{\mu}^3 + c_W\left(\sqrt{3}\ t_W\ W^8_{\mu}
+\sqrt{1- 3\ t^2_W}\  B_{\mu}\right),\nonumber\\
Z_\mu  &=& c_W  W_{\mu}^3 - s_W\left(\sqrt{3}\ t_W\ W^8_{\mu}
+\sqrt{1- 3\ t^2_W}\  B_{\mu}\right),\nonumber\\
Z'_\mu &=&-\sqrt{1- 3\ t^2_W}\ \ W^8_{\mu} + \sqrt{3}\ t_W\
B_{\mu}, \label{apstat} \eea where we use the following notations:
$c_W \equiv \cos \theta_W, s_W \equiv \sin \theta_W $ and $t_W
\equiv \tan \theta_W$.
The {\it physical} states are a mixture of
$Z$ and $Z'$:
\bea
Z_1  &=&Z\cos\phi - Z'\sin\phi,\nonumber\\
Z_2  &=&Z\sin\phi + Z'\cos\phi,\nonumber
\eea
where $\phi$ is a
mixing angle.

\hs  Symmetry breaking and fermion mass generation can be achieved
by three scalar  $\mbox{SU(3)}_L$ triplets $\Phi, \Delta, \Delta'$
and a sextet $\eta $ \bea \Phi & = &\left(
\begin{array}{c}
                \phi^{++}\\ \phi^+\\ \phi^{,0}\\
                \end{array}  \right) \sim (1, 3, 1),\nonumber \\
\label{mh1} \Delta & =& \left( \begin{array}{c}
                \Delta^+_1\\ \Delta^0\\ \Delta^-_2\\
                \end{array}  \right) \sim (1, 3, 0),\nonumber \\
\Delta' & =& \left( \begin{array}{c}
                \Delta^{'0}\\ \Delta^{'-}\\ \Delta^{'--}\\
                \end{array}  \right) \sim (1, 3, -1),\nonumber \\
\eta & = & \left( \begin{array}{ccc}
\eta^{++}_1 & \eta^+_1/ \sqrt{2} & \eta^0/ \sqrt{2}\\
\eta^+_1/ \sqrt{2} & \eta^{'0} & \eta^-_2/ \sqrt{2}\\
\eta^0/ \sqrt{2} & \eta^-_2/ \sqrt{2} & \eta^{--}_2
                \end{array}  \right) \sim (1, 6, 0).\nonumber
\eea The sextet $\eta$ is necessary to give masses to charged
leptons~\cite{fhft,dng}. The vacuum expectation value (VEV)
$\langle \Phi^T \rangle = ( 0, 0, u/ \sqrt{2})$ yields masses for
the exotic quarks, the heavy neutral gauge boson ($Z'$) and two
new charged gauge bosons ($X^{++}, Y^+$). The masses of the
standard gauge bosons and the ordinary fermions are related to the
VEVs of the other scalar fields, $\langle \Delta^0 \rangle = v/
\sqrt{2}, \langle \Delta'^0 \rangle = v'/ \sqrt{2}$ and $\langle
\eta^0 \rangle = \omega/ \sqrt{2},\ \langle \eta^{'0} \rangle =0
$. In order to be consistent with the low energy phenomenology the
mass scale of $SU(3)_L\otimes U(1)_N$ breaking has to be much
larger than that of the electroweak scale, i.e, $u \gg\ v,\ v',
\omega$. The masses of gauge bosons are explicitly given by \be
m^2_W=\frac{1}{4}g^2(v^2+v^{'2}+\omega^2),\
M^2_Y=\frac{1}{4}g^2(u^2+v^2+\omega^2),
M^2_X=\frac{1}{4}g^2(u^2+v^{'2}+4 \omega^2), \label{mnhb} \ee and
\bea m_{Z}^2   &=&\frac{g^2}{4 c_W^2}(v^2+v^{'2}+\omega^2)=
\frac{m_W^2}{c_W^2},\nonumber \\
M_{Z'}^2 &=&\frac{g^2}{3}\left[\frac{c^2_W}{1 - 4 s^2_W} u^2 +
\frac{1 - 4 s^2_W}{4 c^2_W}( v^2 + v^{'2} + \omega^2 ) + \frac{3
s^2_W}{1 - 4 s^2_W} v^{'2}\right]. \label{masmat} \eea

\hs  Expressions in (\ref{mnhb}) yield a splitting between the
bilepton masses~\cite{lng} \be | M_X^2 - M_Y^2 | \leq 3\  m_W^2.
\label{maship} \ee

\hs  Combining constraints from direct searches and neutral
currents, one obtains a range for the mixing angle~\cite{dng} as
$- 1.6 \times 10^{-2} \le \phi \le 7 \times 10^{-4}$ and a lower
bound on $M_{Z_2}: M_{Z_2}\ge 1.3 $ TeV. Such a small mixing angle
can safely be neglected. In that case, $Z_1$ and $Z_2$ are the $Z$
boson  in the SM and the extra $Z'$ gauge boson, respectively.
With the new atomic parity violation in cesium, one gets a lower
bound for the $Z_2$ mass~\cite{ltrun}: $M_{Z_2} > 1.2 $ TeV.

\subsection{The model with RH neutrinos}
\hs In this model the leptons are in triplets, and the third
member is a RH neutrino~\cite{flt,hnl}: \be f_{aL} = \left(
\begin{array}{c}
               \nu_{aL}\\ e_{aL}\\ (\nu^c_L)_a
\end{array}  \right) \sim (1, 3, -1/3), e_{aR}\sim (1,
1, -1). \label{l2} \ee

The first two generations of quarks are in antitriplets while the
third one is in a triplet: \be Q_{iL} = \left( \begin{array}{c}
                d_{iL}\\-u_{iL}\\ D_{iL}\\
                \end{array}  \right) \sim (3, \bar{3}, 0),
\label{q} \ee
\[ u_{iR}\sim (3, 1, 2/3), d_{iR}\sim (3, 1, -1/3),
D_{iR}\sim (3, 1, -1/3),\ i=1,2,\] \be
 Q_{3L} = \left( \begin{array}{c}
                 u_{3L}\\ d_{3L}\\ T_{L}
                \end{array}  \right) \sim (3, 3, 1/3),
\ee
\[ u_{3R}\sim (3, 1, 2/3), d_{3R}\sim (3, 1, -1/3), T_{R}
\sim (3, 1, 2/3).\] The doubly charged bileptons of the minimal
model are replaced here by complex neutral  ones as follows \bea
\sqrt{2}\ W^+_\mu &=& W^1_\mu - iW^2_\mu ,
\sqrt{2}\ Y^-_\mu = W^6_\mu - iW^7_\mu ,\nonumber\\
\sqrt{2}\ X_\mu^o &=& W^4_\mu - iW^5_\mu. \eea

\hs  The {\it physical} neutral gauge bosons are again related to
$Z, Z'$ through the mixing angle $\phi$. Together with the photon,
they are defined as follows~\cite{hnl}
\begin{eqnarray}
A_\mu  &=& s_W  W_{\mu}^3 + c_W\left(- \frac{t_W}{\sqrt{3}}\
W^8_{\mu} +\sqrt{1-\frac{t^2_W}{3}}\  B_{\mu}\right),
\nonumber\\
Z_\mu  &=&  c_W  W^3_{\mu} - s_W\left( -\frac{t_W}{\sqrt{3}}\
W^8_{\mu}+
\sqrt{1-\frac{t_W^2}{3}}\  B_{\mu}\right),  \\
Z'_\mu &=& \sqrt{1-\frac{t_W^2}{3}}\
W^8_{\mu}+\frac{t_W}{\sqrt{3}}\ B_{\mu}.\nonumber \label{apstat1}
\end{eqnarray}
The symmetry breaking can be achieved with just three
$\mbox{SU}(3)_L$ triplets  \bea \chi & = &\left(
\begin{array}{c}
                \chi^0\\ \chi^-\\ \chi^{,0}\\
                \end{array}  \right) \sim (1, 3, -1/3),\\
\label{h1} \rho & =& \left( \begin{array}{c}
                \rho^+\\ \rho^0\\ \rho^{,+}\\
                \end{array}  \right) \sim (1, 3, 2/3),\\
\eta & =& \left( \begin{array}{c}
                \eta^0\\ \eta^-\\ \eta^{,0}\\
                \end{array}  \right) \sim (1, 3, -1/3).\\
\eea The necessary VEVs are \be \langle\chi \rangle^T = (0, 0,
\omega/\sqrt{2}),\ \langle\rho \rangle^T = (0, u/\sqrt{2}, 0),\
\langle\eta \rangle^T = (v/\sqrt{2}, 0, 0). \label{vev} \ee

 The VEV $\langle \chi \rangle$ generates masses for the exotic 2/3 and
--1/3 quarks, while the  VEVs $\langle \rho \rangle$ and $\langle
\eta \rangle$ generate masses for all ordinary leptons and quarks.
 After symmetry breaking
the gauge bosons gain  masses as \be
m^2_W=\frac{1}{4}g^2(u^2+v^2),\
M^2_Y=\frac{1}{4}g^2(v^2+\omega^2),
M^2_X=\frac{1}{4}g^2(u^2+\omega^2), \label{rhb} \ee and \bea
m_{Z}^2   &=&\frac{g^2}{4 c_W^2}(u^2+v^2)=
\frac{m_W^2}{c_W^2},\nonumber \\
M_{Z'}^2 &=&\frac{g^2}{4(3-4s_W^2)}\left[4 \omega^2+
\frac{u^2}{c_W^2} + \frac{v^2(1-2s_W^2)^2}{c_W^2}\right].
\label{masmat} \eea

\hs  In order to be consistent with the low energy phenomenology
we have to assume that $\langle \chi \rangle \gg\ \langle \rho
\rangle,\ \langle \eta \rangle$ such that $m_W \ll M_X, M_Y$.

\hs  The symmetry-breaking hierarchy gives us a splitting between
the bilepton masses~\cite{til} \be | M_X^2 - M_Y^2 | \leq m_W^2.
\label{mashipr} \ee Therefore it is acceptable to put $M_X \simeq
M_Y$.

\hs  The constraint on the $Z - Z'$ mixing based on the $Z$ decay
is given~\cite{hnl}: $-2.8 \times 10^{-3} \le \phi \le 1.8 \times
10^{-4}$, and in this model we have not  a limit for $\sin^2
\theta_W$. With so  small mixing angle, $Z_1$ and $Z_2$ are the
$Z$ boson in the SM  and the extra $Z'$ gauge boson, respectively.
From the data on parity violation in the cesium atom, one gets a
lower bound on the $Z_2$ mass in range between 1.4 TeV and 2.6
TeV~\cite{ltrun}. Data on the kaon mass difference $\Delta m_K$
gives a bound~\cite{lv}: $M_{Z_2} \leq 1.02$ TeV.

\section{$Z'$ production in $e^-\  \gamma $ collisions }

\hs Now we are interested in the single production of new neutral
gauge bosons $Z'$ in $e^-\   \gamma $ collisions

\begin{equation}
 e^{-}(p_{1},\lambda) + \gamma(p_{2},\lambda^{'})  \rightarrow
  e^{-}(k_{1},\tau) + Z^{'}(k_{2},\tau^{'}),
  \label{proc1}
\end{equation}
where $p_i$, $k_i$ stand for the momenta and  $\lambda$,
$\lambda'$, $\tau$, $\tau'$ are the  helicities of the particles.
At the tree level, there are two Feynman diagrams contributing to
the reaction (\ref{proc1}),
depicted in Fig. 1 \vspace*{0.2cm}

\begin{center}
\begin{picture}(300,50)(-20,0)
\Photon(-10,35)(5,10){2}{4} \ArrowLine(-10,-15)(5,10)
\ArrowLine(5,10)(55,10) \Photon(55,10)(70,35){2}{4}
\ArrowLine(55,10)(70,-15) \Text(32,18)[]{$e^{-} $}
\Text(-16,40)[]{$\gamma$} \Text(-16,-20)[]{$e^-$}
\Text(80,40)[]{$Z^{'}$} \Text(80,-20)[]{$e^-$} \Text(185,-42)[]{ }

\Photon(200,35)(250,10){2}{6}
 \Photon(200,10)(250,35){2}{6}
\ArrowLine(185,-15)(200,10)
 \ArrowLine(200,10)(250,10)
\ArrowLine(250,10)(265,-15)

\Text(230,5)[]{$e^{-} $}
 \Text(195,40)[]{$\gamma$}
\Text(180,-20)[]{$e^-$}
\Text(270,40)[]{$Z^{'}$}
\Text(265,-20)[]{$e^-$}
\end{picture}


\vspace{1cm}
 Figure 1: Feynman diagrams for $\ e^{-} \gamma
\rightarrow  Z' \ e^{-} $
\end{center}
 \vspace*{1cm}

\hs  The $s$ - channel amplitude is given by
\begin{equation}
M_{s}^{Z'}= \frac{ieg}{2c_{W}q^2_s}
\epsilon_{\mu}(p_{2})\epsilon_{\nu}(k_{2})
\overline{u}(k_{1})\gamma^{\nu}[g_{2V}(e)-g_{2A}(e)\gamma_{5}]
q\!\!\!/_s\gamma^{\mu}u(p_{1}),
\end{equation}
where $q_s= p_1+p_2 $. The $u $ - channel amplitude is
\begin{equation}
M_{u}^{Z'}= \frac{ieg}{2c_{W}q^2_u}
\epsilon_{\mu}(k_{2})\epsilon_{\nu}(p_{2})
\overline{u}(k_{1})\gamma^{\nu}q\!\!\!/_u\gamma^{\mu}
[g_{2V}(e)-g_{2A}(e)\gamma_{5}] u(p_{1}),
\end{equation}
here  $q_u= p_1-k_2 $ and  $\epsilon_{\mu}(p_{2})$,
$\epsilon_{\nu}(p_{2})$ and $\epsilon_{\nu}(k_{2})$,
$\epsilon_{\mu}(k_{2})$ are the polarization vectors of the photon
$\gamma$ and the $Z^{'}$ boson, respectively, $ g_{2V}(e),
g_{2A}(e) $ are coupling constants of $Z'$ to the electron $e$. In
the minimal model they are given by~\cite{dng} \be g_{2V}(e)=
\frac{\sqrt{3}}{2}\sqrt{1- 4 s_W^2} , \hs g_{2A}(e)=-
\frac{1}{2\sqrt{3}}\sqrt{1- 4 s_W^2}, \label{hsttm}\ee while in
the model with RH neutrinos~\cite{hnl}
 \be g_{2V}(e)=
\left(-\frac 1 2 + 2 s_W^2\right)\frac{1}{\sqrt{3-4 s_W^2}}, \hs
g_{2A}(e)= \frac{1}{2\sqrt{3-4 s_W^2}}. \label{hsttr}\ee From Eqs.
(\ref{hsttm}) and (\ref{hsttr}) we see that due to the factor
$\sqrt{1- 4 s_W^2}\ll 1 $, the cross sections in the minimal model are
smaller than that in the model with RH neutrinos.
 We  work in the center-of-mass frame and
denote the scattering angle (the angle  between momenta of the
initial electron and the final one) by  $\theta$. We have
evaluated the $\theta$ dependence of the differential
cross-section $d\sigma/d \cos \theta$, the energy and the
$Z'$ boson mass dependence of the total cross-section $\sigma$.\\

\hs i) In Fig. 2 we plot $d \sigma/d cos{\theta}$ for the minimal
model as a function of $cos{\theta}$ for the collision energy at
CLIC $\sqrt{s}=2733 $ GeV~\cite{rwa} and the relatively low value
of mass $m_{Z'}=800$ GeV. From Fig. 2 we see that $d \sigma/d
cos{\theta}$ is peaked in the backward direction (this is due to
the $ e^-$ pole term in the $u$-channel) but it is flat in the
forward direction. Note that the behaviour of $d \sigma/d
cos{\theta}$ for the model with RH neutrinos  is similar at other
values of $\sqrt{s}$.

\hs ii) The energy dependence of the cross-section for the minimal
model is shown in Fig. 3. The same value of the mass as in i),
$m_{Z'}=800$ GeV is chosen. The energy range is 1200 GeV $\leq$
$\sqrt{s}$ $\leq$ 3000 GeV. The curve (a) is the total
cross-section for the minimal model, the curves (b) and (c)
represent cross-sections of the $u, s$-channel only, respectively.
The curve (d) is the cross-section for the SM model, reduced three
times. The $u$-channel, curve (b) rapidly decreases with
$\sqrt{s}$ while the $s$-channel has a zero point at
$\sqrt{s}=m_{Z'}$ then slowly increases. In the high energies
limit, the $s$-channel gives main contribution to the total
cross-section. In Fig. 3 the cross-section of the standard model
reaches 0.18 pb then slowly decreases to  0.05 pb while the
cross-section of the minimal model is only  0.14 pb at
$\sqrt{s}=800$  GeV and  0.05 pb at $\sqrt{s}=2733 $  GeV. The
same situation also occurs in the model with RH neutrinos. In this
model we fix $m_{Z'}=800 $  GeV and illustrate the energy
dependence of the cross-section in Fig. 4. The energy range is the
same as Fig. 3, $ 1200$  GeV $\leq \sqrt{s} \leq $  3000 GeV. We
see from Fig. 4 that the cross-section $\sigma$ decreases with
$\sqrt{s}$, from $\sigma$=0.35 pb down to  $\sigma$=0.08 pb.\\

\hs iii) We have plotted the boson mass dependence of the number
of events in three models in Fig. 5. The energy is fixed
$\sqrt{s}=2733  $  GeV and the mass range is $800$  GeV $\leq
m_{Z'}\leq 2000$ GeV. As we mentioned above, due to coupling
constant, the order of the line of number of events, from bottom
to top, is minimal, SM, model with RH neutrinos. The smallest
number of events is of the minimal model. With the integrated
luminosity $L \simeq  100 fb^{-1}$, the number of events can be
several thousands.

\hs In final state $Z^{'}$ will decay into leptons and quarks. Its
partial decay width equals~\cite{pdg}
\[
\Gamma(Z^{'} \to f \bar{f})=
\frac{G_{F}m_{Z^{'}}^{2}}{6\sqrt{2}\pi}
N_{c}^{F}[(g_{2A}^{f})^{2}R_{A}^{f} + (g_{2V}^{f})^{2}R_{V}^{f}]
 = \left \{ \begin{array}{ll}
6.4 & \mbox{GeV for minimal model}\\
11.8 & \mbox{GeV for RH neutrinos model.}
\end{array}
\right.\] Due to coupling constants, the lifetime of $Z^{'}$ in
the minimal model is longer than that in the model with RH
neutrinos.

\section{Conclusion }
\hs In this paper, we have presented  the production of single
$Z^{'}$ boson in the $e^- \gamma$ reaction in the framework of the
3 - 3 - 1 models. We see that with this process, the reaction
mainly occurs at small scattering angles. The results show that if
the mass of the boson is in a range of 800 GeV, then single boson
production in $e^- \gamma$ collisions may give observable value at
moderately high energies. At CLIC based on $e^- \gamma$ colliders,
with the integrated luminosity $L \simeq  100 {fb}^{-1} $ one
expects observable experiments in future colliders. Due to the
values of the coupling constants,  cross sections in the model
with RH neutrinos are bigger than those in the minimal one.

\hs In conclusion, we have pointed out the usefulness of electron
- photon colliders in testing the 3 - 3 - 1 models at high
energies, through the reaction $e^- \gamma \rightarrow e^- Z'$. If
the $Z'$ boson is not so heavy, this reaction offers a much better
discovery reach for $Z'$ than the pair production in $e^+ e^-$ or
$e^- e^-$ collisions.

\hs {\it Acknowledgments}: One of the authors (H. N. L) would like
to thank Members of Physics Division, National Center for
Theoretical Sciences (NCTS), Hsinchu, Taiwan, where this work was
completed, for warm hospitality during his visit. His work is supported
by the NCTS under grant from National Science Council
of Taiwan R. O. C. This work was supported in
part by the National Council for
Natural Sciences of Vietnam.\\

 \setcounter{figure}{1}
\begin{figure}[b]
\begin{center}
\includegraphics[width=10cm,height=7cm]{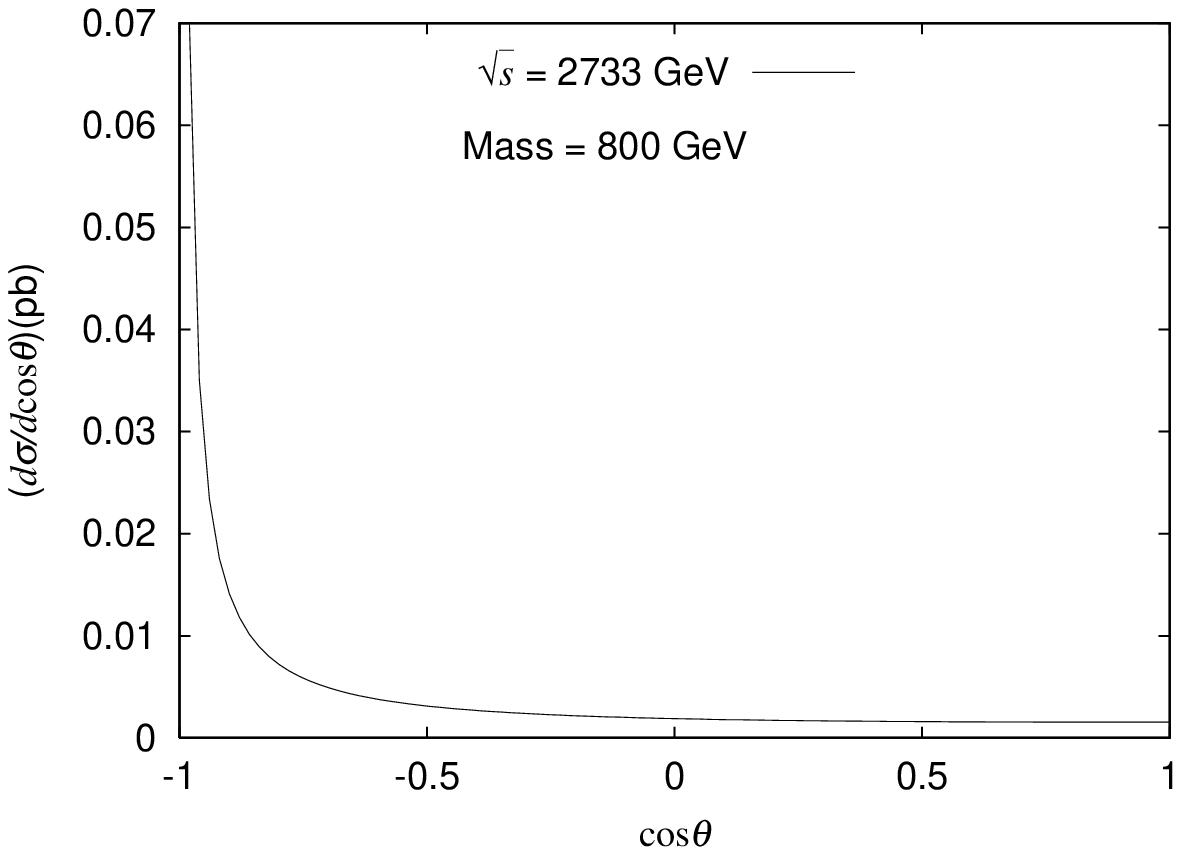}
\caption{\label{lim}{\em Differential cross section of the minimal
model  }}
\end{center}
\end{figure}
\vspace{2cm}

 \setcounter{figure}{2}
\begin{figure}[b]
\begin{center}
\includegraphics[width=10cm,height=7cm]{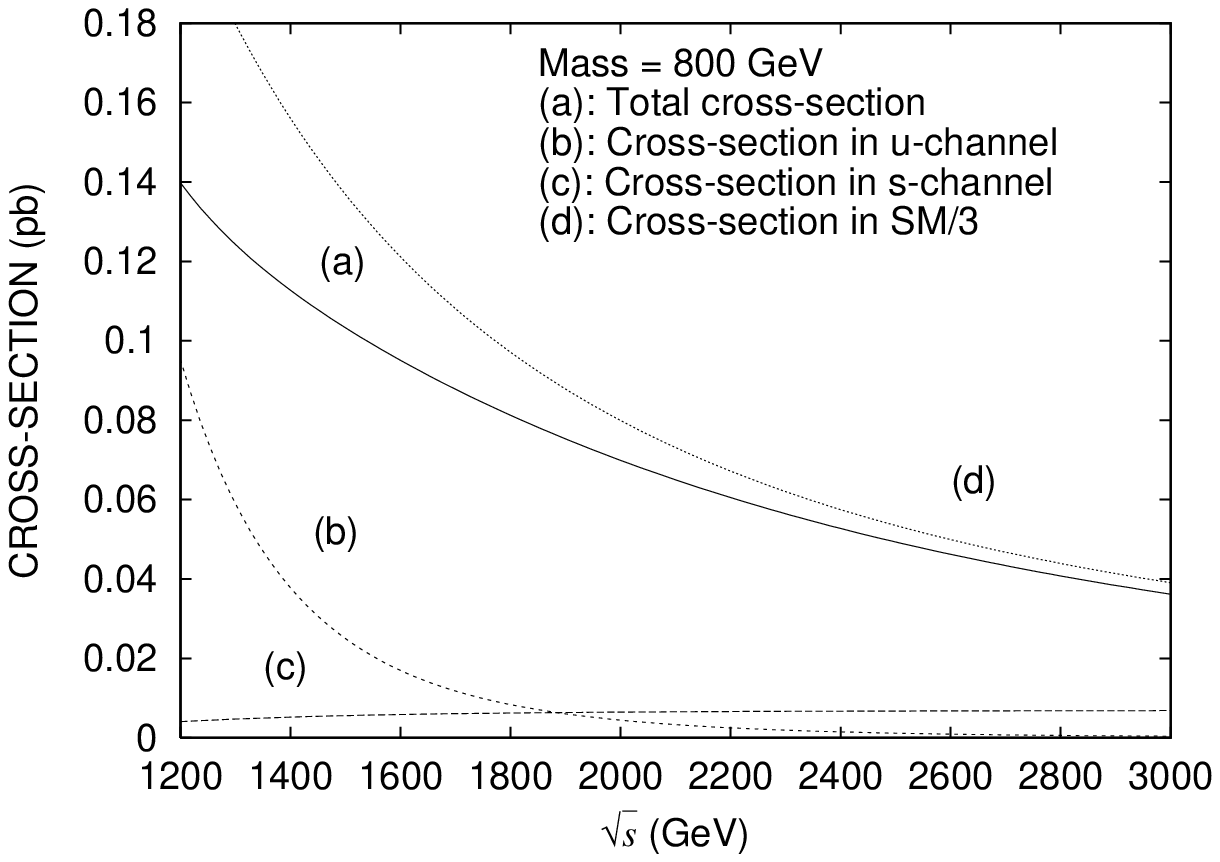}
\caption{\label{lim}{\em Cross section $\sigma (e \gamma
\rightarrow Z' e)$ of the minimal  model as a function of
$\sqrt{s}$}}
\end{center}
\end{figure}
\vspace{2cm}

 \setcounter{figure}{3}
\begin{figure}[b]
\begin{center}
\includegraphics[width=10cm,height=7cm]{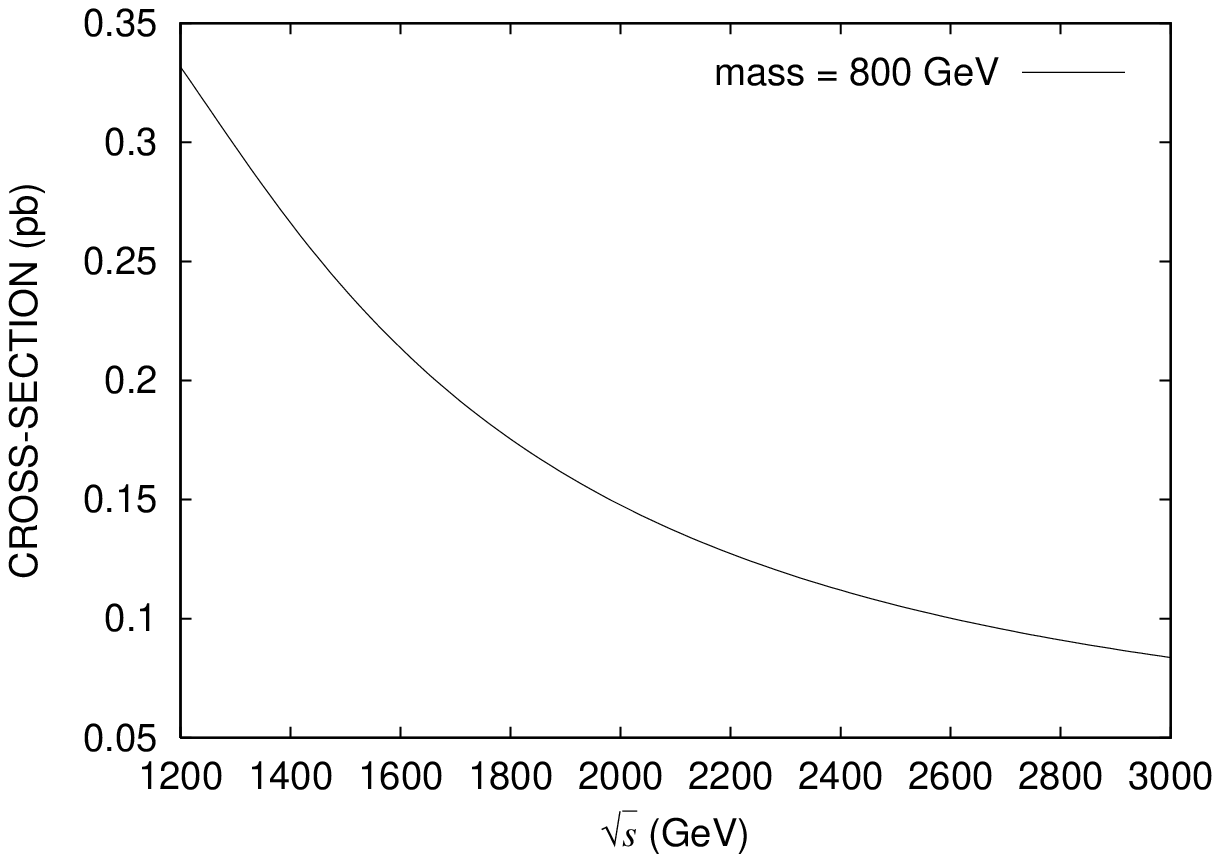}
\caption{\label{lim}{\em Cross section $\sigma (e \gamma
\rightarrow Z' e)$ of the model with RH neutrinos   as a function
of $\sqrt{s}$}}
\end{center}
\end{figure}
\vspace{2cm}

\setcounter{figure}{4}
\begin{figure}[b]
\begin{center}
\includegraphics[width=10cm,height=7cm]{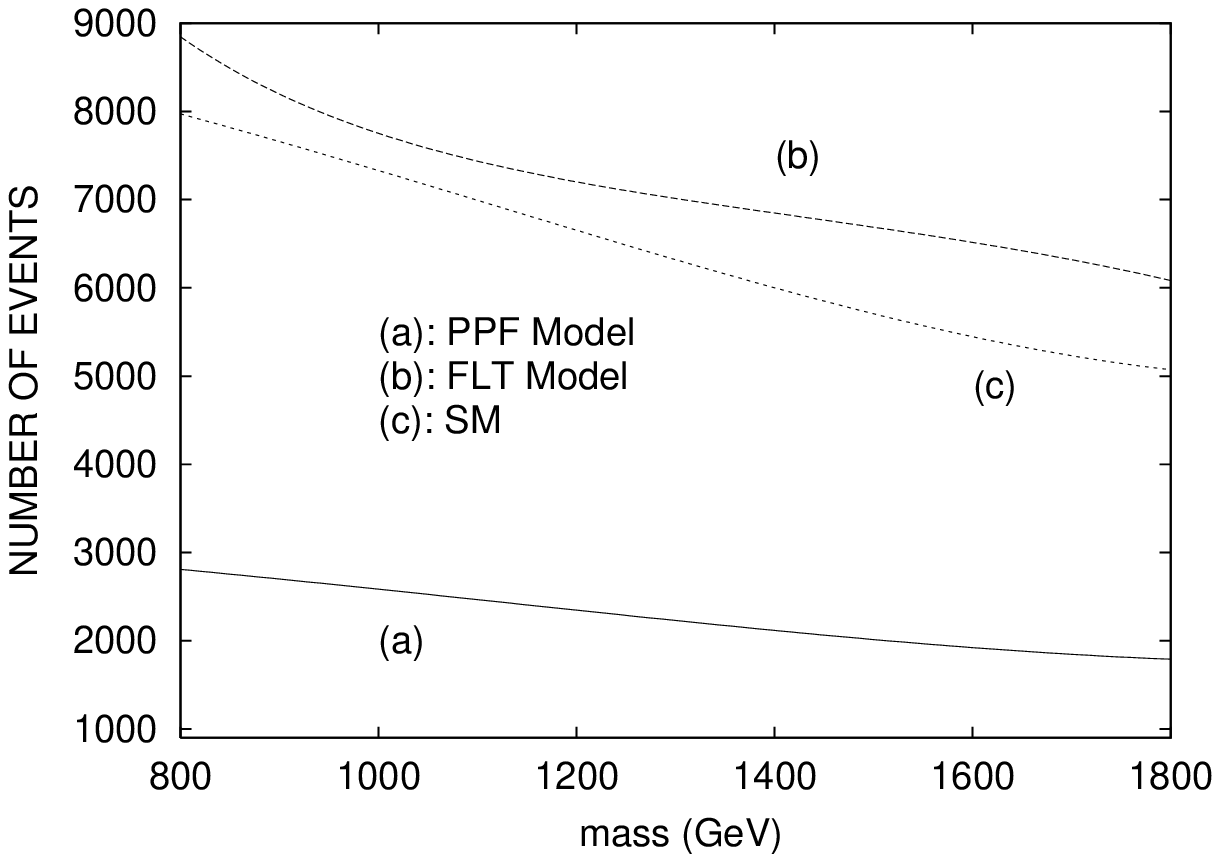}
\caption{\label{lim}{\em Number of events of three models}}
\end{center}
\end{figure}

\end{document}